\documentclass{aa}  
\usepackage{graphicx}
\usepackage{txfonts}
\begin{document}
\title{Spatially resolved physical and chemical properties of the
  planetary nebula NGC\,3242\thanks{ Based on observations collected
    at the European Southern Observatory, Chile, via programme ESO
    No. 074.D-0425(A)}}

\subtitle{} 

  \author{H. Monteiro \inst{1}, D. R. Gon\c calves \inst{2},
    M. L. Leal-Ferreira \inst{3}, R.L.M. Corradi \inst{4,5} }

          \institute{Departamento de F\'isica e Qu\'imica,
            Universidade Federal de Itajub\'a. \email{hmonteiro@unifei.edu.br} 
	 \and 
            Observat\'orio do Valongo, Universidade Federal do Rio de Janeiro, 
            Ladeira Pedro Antonio 43, 20080-090 Rio de Janeiro, Brazil.    
	 \and 
	    Argelander-Institut f\"ur Astronomie, University of Bonn, Auf dem H\"ugel 71,
            D-53121 Bonn, Germany. 
            \and 
            Instituto de Astrof\'\i sica de Canarias, E-38200 La Laguna, 
            Tenerife, Spain
            \and
Departamento de Astrof\'\i sica, Universidad de La Laguna, 
E-38206 La Laguna, Tenerife, Spain}

\titlerunning{Spatially resolved properties of NGC~3242} 

\authorrunning{Monteiro et al.} 

 \date{Received ; accepted }

 
  \abstract
  { } {Optical integral-field spectroscopy was used to investigate the
    planetary nebula \object{NGC~3242}. We analysed the main
    morphological components of this source, including its knots, but not 
    the halo. In addition to revealing the properties of
    the physical and chemical nature of this nebula, we also
    provided reliable spatially resolved constraints that can be used
    for future photoionisation modelling of the nebula. The latter is
    ultimately necessary to obtain a fully self-consistent
    3D picture of the physical and chemical properties of the
    object.}  {The observations were obtained with the VIMOS
    instrument attached to VLT-UT3. Maps and values for specific
    morphological zones for the detected emission-lines were obtained
    and analysed with routines developed by the authors to derive
    physical and chemical conditions of the ionised gas in a 2D
    fashion.}  { We obtained spatially resolved maps and mean values
    of the electron densities, temperatures, and chemical abundances,
    for specific morphological structures in NGC~3242. These results
    show the pixel-to-pixel variations of the the small- and
    large-scale structures of the source. These diagnostic maps
    provide information free from the biases introduced by traditional
    single long-slit observations.}  { In general, our results are
    consistent with a uniform abundance distribution for the object,
    whether we look at abundance maps or integrated fluxes from
    specified morphological structures.  The results indicate that
    special care should be taken with the calibration of the data and
    that only data with extremely good signal-to-noise ratio and
    spectral coverage should be used to ensure the detection of
    possible spatial variations. }


   \maketitle
%

\section{Introduction}

Planetary nebulae (PNe) are end products of the evolution of stars
 with masses from approximately 0.8 to 8 M$_{\odot}$ and as such
have great importance in many fields in astrophysics, from basic
atomic processes in solar-like stars to distant galaxies. Although the
general picture of planetary nebulae formation is well understood
(\cite{K08}), many questions remain unsolved, such as the mechanism by
which the material ejected by the star finally forms the many
observed morphologies (\cite{balickfrank02}).

NGC~3242 is a multiple-shell PN within a bright 28''$\times$20''
inner elliptical shell, which also contains a pair of ansae 
(or, small-scale low-ionisation
structure, LIS; \cite{goncalves01}). The inner shell is surrounded by a
fainter 46''$\times$40'' moderately elliptical envelope (see, for
instance, \cite{ruiz11}). These two shells are further more enclosed by
concentric rings and a giant broken halo revealed by deep images
(Corradi et al. 2004; \cite{monreal05}).

The bright central star (V=12.43) of the nebula has been studied by
several authors. Different methods for evaluating its effective
temperature (T$_{eff}$) were applied, which thus culminates in a wide
range of possible values. In a detailed discussion, \cite{pottasch08}
showed that the hydrogen Zanstra temperature is 57,000~K, while the
helium Zanstra temperature is 90,000~K. From stellar atmospheric
models 
the T$_{eff}$ is found to be 75,000~K (\cite{pauldrach04}), but
temperatures as high as 94,000~K are found in the literature
(\cite{tinkler02}). After analysing this wide range of possibilities and 
taking into account the higher helium Zanstra temperature, \cite{pottasch08}
assumed in their work a temperature of 80,000~K for the central star
of NGC~3242, which leads to a luminosity of 730~L$_\odot$, with a
radius of 0.141 R$_\odot$. Notice, however, that other authors have
found higher values for the luminosity of this nebula. For example,
the 3200~L$_\odot$ found by Pauldrach et al. (2004). The mass-loss
rate of the central star is estimated to be
$\leq$2$\times$10$^{-8}$M$_{\odot}$/yr (\cite{kudritzki97}) and the
post-AGB mass is believed to lie between 0.53~M$_{\odot}$ and
0.56~M$_{\odot}$. The latter value corresponds to an MS star of
1.2$\pm$0.2~M$_{\odot}$
(Galli et al. 1997; Stanghellini \& Pasquali 1995; \cite{pauldrach04}).

For the nebula in the radio wavelength --its optical
properties are discussed in details below--, the
H91$\alpha$ recombination line and the 3.5cm continuum observations
(Rodr\' iguez et al. 2010) give nebular temperatures
(10,100$\pm$700~K) consistent, within 10\%, with that obtained from
optical lines and the Balmer discontinuity (\cite{liu93};
\cite{krabble06}). Moreover, that the LTE continuum and the line
H91$\alpha$ temperatures agree very well is in proof itself 
that a significant continuum contamination by dust emission is absent 
from this nebula. Still following \cite{rodriguez10}, the radio
continuum measurements can be interpreted as being completely due to
free-free emission, without the 30~GHz excess previously suggested by
\cite{casassus07}.

The observations of \cite{ruiz11}, taken with the XMM-Newton, have shown
that the X-ray luminosity of NGC~3242 is $\sim$
2$\times$10$^{30}$~erg~s$^{-1}$ for an adopted distance of 0.55~kpc
(Terzian 1997; Mellema 2004). Following the same authors, the
temperature of the hot-bubble region in which the X-rays are
produced is $\sim$2.35$\times$10$^6$~K.  These figures agree with the 
ad hoc predictions of the shock-heated stellar wind models
confined by heat conduction.

Few works address the question of the spatial variation of the nebular
properties, and if they do, they use traditional long-slit
data. One example is the study of \cite{pc98} about 20 PNe (mostly
Type I), showing that PNe are in general chemically homogeneous. 
\cite{balick94} reported that a few objects (specifically, NGC~6543,
6826, and 7009) showed significant abundance variations from one
component to another, but subsequent 3D photoionisation modelling did not
confirm it (Gon\c calves et al. 2006). Although NGC 3242 is a well-known 
PN, with about 350 references
from 1990 to date, and even though different diagnostics were clearly 
determined, only few works investigated its characteristics using some kind 
of spatially resolved data, from a range of wavelengths. \cite{monreal05}, 
the only work apart from ours that used IFU data for this object, succeeded 
in determining some properties of the halo, but did not perform a detailed 
analysis of the physical and chemical properties.

Our aim here is to derive the ionic and total element abundances 
for NGC~3242 in a spatially resolved way from VLT VIMOS IFU data. We aim 
to address the question of the chemical abundance variation within the PN. 
Our motivation is that, in principle, material at different positions 
within a PN (or different PN components) may be the result of distinct 
mass-loss episodes of the progenitor star and therefore may trace the chemical
inhomogeneities (enrichment) in the original (not coeval) outflows.

\begin{figure}[!ht]
\begin{center}
\includegraphics[width=\columnwidth]{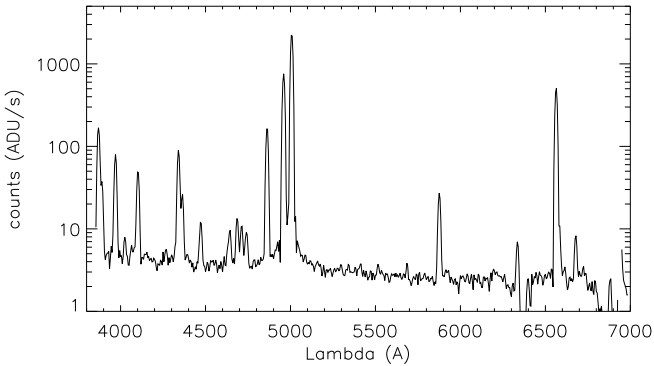}\\
\caption{Spectrum from the spaxel (50,30) showing typical features and
  quality of the data.}
\label{spaxel}
\end{center}
\end{figure}

In Section 2 we discuss the data and their reduction, while in Section 3 we
show the results obtained from the emission maps. In Section 4 we
present the results for specific morphological structures on which
fluxes were integrated, and in Section 5 we conclude.

\section{Data acquisition and reduction} 

\begin{figure}[!ht]
\begin{center}
\includegraphics[width=\columnwidth]{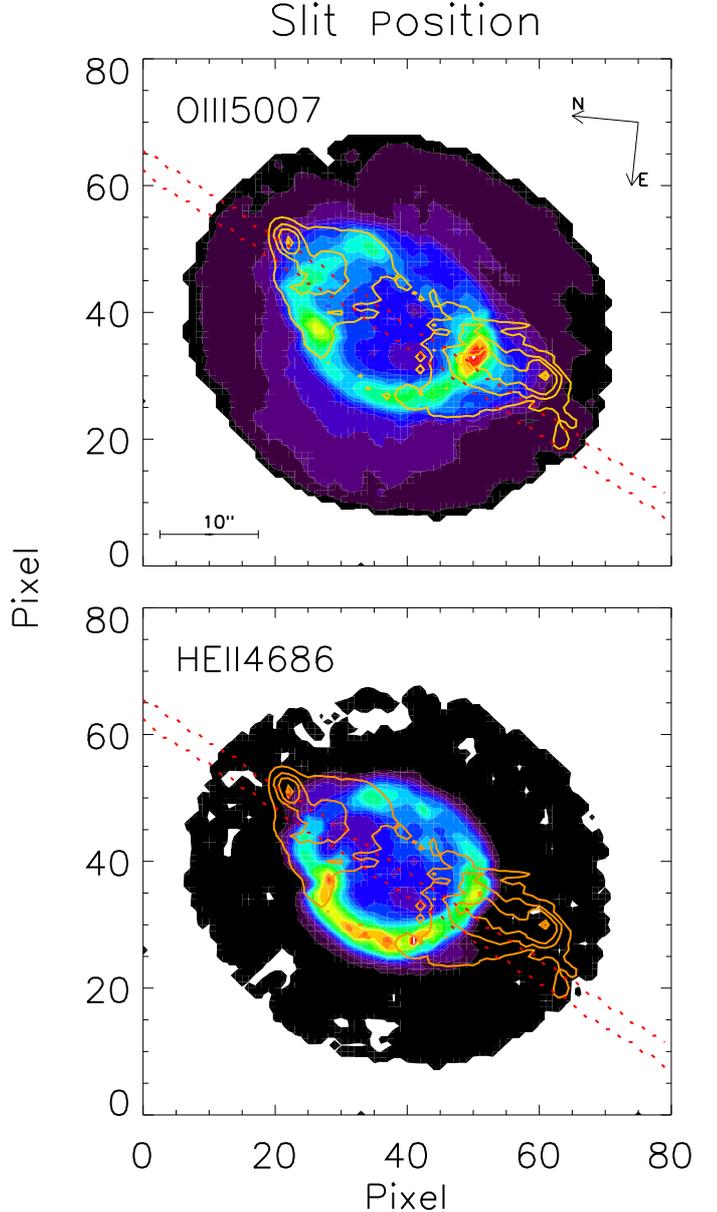}\\
\caption{Line-emission maps for [O~{\sc iii}]~$\lambda$5007 (top) and
  He{\sc ii}~$\lambda$4686 (bottom) showing the position of the
  long-slit observations used for the flux calibration of the data (red
  dotted lines). The orientation in the sky as well as the plate scale
  is also indicated in the top map. The contours overlaid in both
  figures are of the [N~{\sc ii}]6584 line-emission map. In the maps
  above the intensity scale is normalised to the maximum value for
  each emission line.}
\label{lna-slit}
\end{center}
\end{figure}

The observations were obtained with the instrument VIMOS-IFU, attached
to VLT-UT3. The instrument is composed of 6400 fibers and, has a
changeable scale on the sky that was set to 0.67$"$ per fiber, to 
obtain our data. The image is formed by a matrix of 80x80 fibers,
which gave us a coverage of 54$"$x54$"$ on sky. We obtained
observations in low-resolution mode, with a pixel scale of
5~\AA~pix$^{-1}$ with a spectral coverage of 3290~\AA, yielding a 
useable range from 3900~\AA\ to 7000~\AA, and in high-resolution
mode with a pixel scale of 0.6~\AA~pix$^{-1}$, yielding a spectral
coverage of 2200~\AA, from 5250~\AA\ to 7450~\AA. The reduction was
performed with the VIMOS pipelines available at the instrument website
\footnote{http://www.eso.org/sci/facilities/paranal/instruments/vimos/}. A
  typical spectrum from the IFU data is shown in Fig. \ref{spaxel}
  where the data for spaxel (50,30) are displayed.

Owing to poor weather we were unable to observe a standard star. To
overcome this limitation, we opted to flux-calibrate the data with
respect to a long-slit spectrum of the object, obtained at a posterior
date, and kindly provided by R. Costa, as described below.

Because we had a resolved emission map of the nebula for all detected
emission lines, the procedure to perform flux calibration was essentially 
the comparison of pixels in the emission maps obtained from the IFU
data with a given observed long-slit configuration. 
In an ideal situation, a given area extracted from the IFU-calibrated 
maps should give the same integrated flux values as an
equivalent aperture or slit from a distinct observation. Clearly, 
atmospheric variations very likely have a significant impact on the
final fluxes because of different seeing and extinction values. However,
because we scaled the emission-line maps to match an integrated flux for
a given long-slit area, at least the relative flux variations within
the maps were preserved.

\begin{figure*}[!ht]
\begin{center}
\includegraphics[scale=0.48]{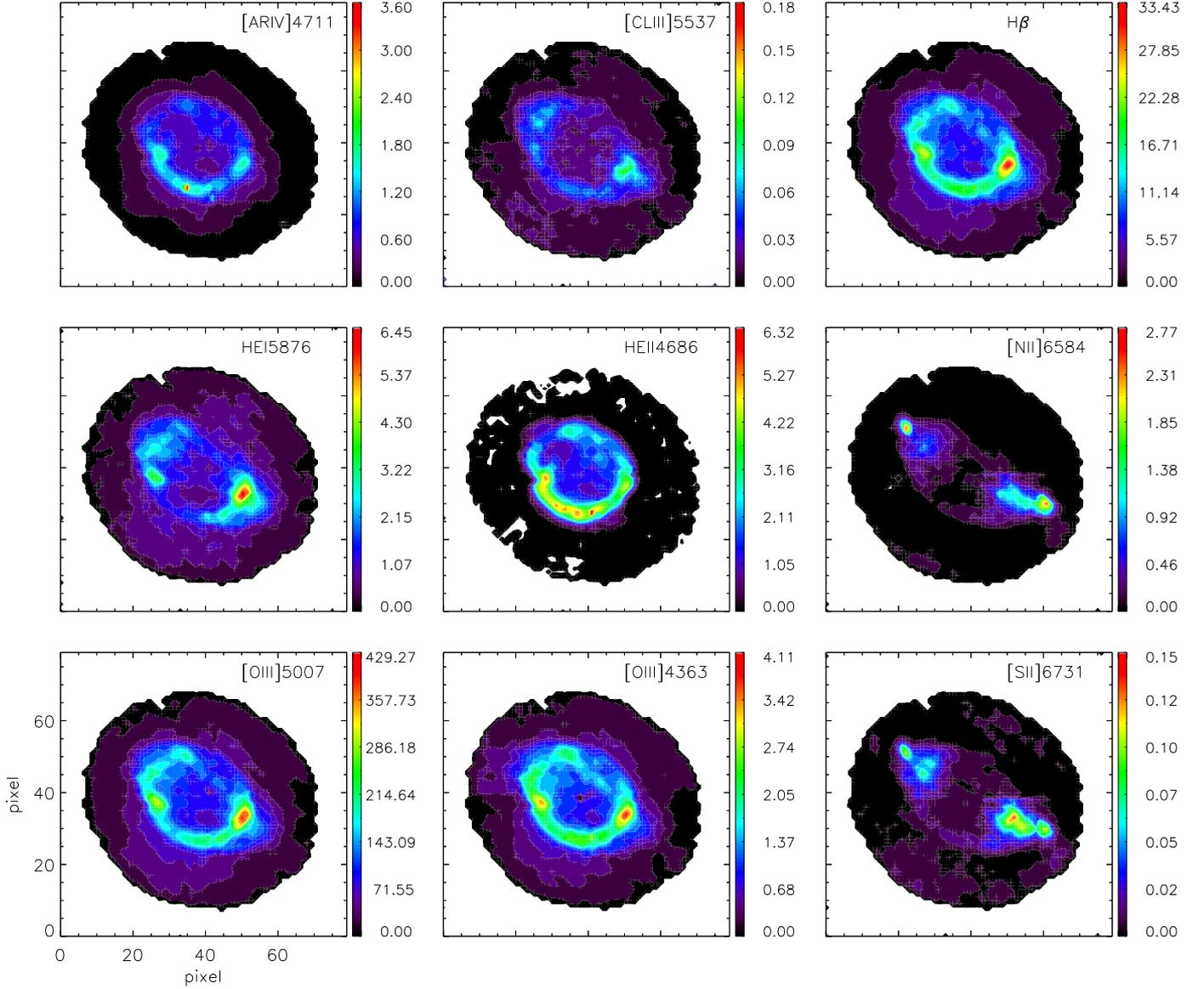}\\
\caption{Emission-line maps for the most important lines observed with
  VIMOS-IFU.  From top to bottom and left to right, they are [Ar~{\sc
    iv}]~4711\AA; [Cl~{\sc iii}]~5537\AA; H$_{\beta}$; He~{\sc
    i}~5876\AA; He~{\sc ii}~4686\AA; [N~{\sc ii}]~6584\AA; [O~{\sc
    iii}]~5007\AA; [O~{\sc iii}]~4363\AA; and [S~{\sc
    ii}]~6731\AA. The orientation in the sky and the plate scale is
  the same as in Fig. 1. The intensity scale is normalised to the
  maximum value for each emission line.}
\label{line-maps}
\end{center}
\end{figure*}

In Figure~1 we show the position of the long-slit used in
the flux calibration of the data as well as emission-line map
information that is discussed in detail below. It is clear from
the figure that the slit position misses the bright parts of both
knots of the nebula. It is also evident that the slit only captures 
the lowest intensity part of the He~{\sc ii}~4686\AA~ emission.

\section{Results} 

In Fig.~\ref{line-maps} we show the emission-line maps for the
stronguest lines we detected, which were used as diagnostic of
the electron densities, temperatures, and chemical abundance. The maps
show a clear sensitivity variation from one lens array to another (see,
for example, the emission-line map of [S~{\sc ii}] in
Fig.~\ref{line-maps}). Although these variations are sometimes not
clear in a given emission-line map, they become more evident in some
diagnostic ratio maps, as discussed below.  There are also dead fibers
in some of the quadrants. These problems are documented in the
instrument pages (VLT-IFU) pages and we refer to them for further
details. In short, we were able to correct for some of the artifacts
generated by these problems, but not all could be removed.

Because we not deal with integrated fluxes, the typical
signal-to-noise ratio in a volumetric pixel (voxel) of the data cube
can be significantly lower in comparison. In practice, the trade-off
of spatial resolution is lower signal to noise ratio (S/N) in a given
pixel of the observed map. For individual emission-line maps the
limited S/N is no great problem, but it results in significant noise
when spatially detailed diagnostic ratio maps are computed.  To
improve the quality of the final maps we applied a fast Fourier filter
to remove some of the noise, especially in the low S/N regions. Areas
with a S/N below 5 were completely removed and their signal was set to
0.

\subsection{Extinction coefficient, densities, and temperatures}

To work with the emission line maps presented in Fig.~\ref{line-maps}
and derive the nebular properties (internal extinction, electron
densities, and temperatures as well as ionic and total chemical
abundances), we used the program {\sc 2d\_neb}. This program is based
on the well established {\sc iraf} {\it nebular} package, and was
developed to enable the use of the spectroscopic maps to easily obtain
the astrophysical quantities of ionised nebulae also in the form of
maps. The {\sc 2d\_neb} package uses the traditional line-ratio
diagnostics and the atomic data from {\sc iraf} to derive N$_e$ and
T$_e$. The process is based on the {\sc iraf.nebular.temden} task and
consists of solving the equation of statistical equilibrium.  The
ionic abundance calculation follows the {\sc iraf.nebular.ionic} task,
complemented by the atomic data given by \cite{benjamin99}. The total
chemical abundances maps are derived using the ICFs given by
\cite{kb94} and \cite{liu00} (see \cite{leal11} for more details on
the {\sc 2d\_neb} code).

The first step in this analysis was the determination of the
extinction coefficient c(H$\beta$) using the Balmer decrement.
Because the H$\alpha$ line was saturated in a number of pixels and the
H$\beta$, H$\gamma$, and H$\delta$ line-ratios only provided noisy
ratio maps, we computed an average c(H$\beta$) value from the Balmer
line fluxes integrated over the whole nebula. We adopted the
extinction curve given by \cite{ccm89}, updated by \cite{odonnell94},
with R$_\nu$=3.1.  We derived c(H${\beta}$)=0.25 (Table~1), which is
consistent with values from the literature, which are around 0.2
(Balick et al. 1993; Henry et al. 2000; Tsamis et al. 2003; Pottasch
\& Bernard-Salas 2008).  We adopted this value to correct all the
VIMOS-IFU emission-line maps we observed for extinction.

\begin{figure}[!ht]
\begin{center}
\includegraphics[width=\columnwidth]{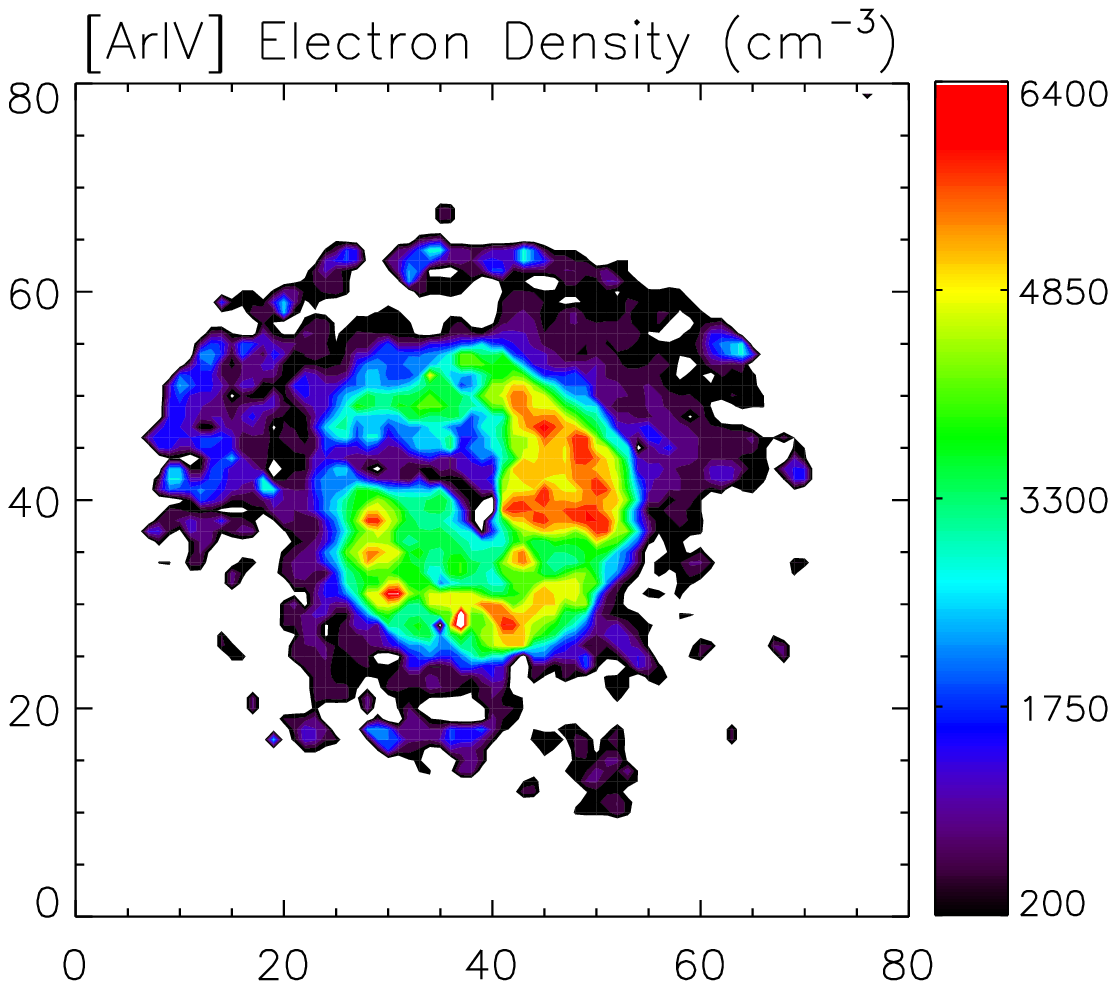}\\
\includegraphics[width=\columnwidth]{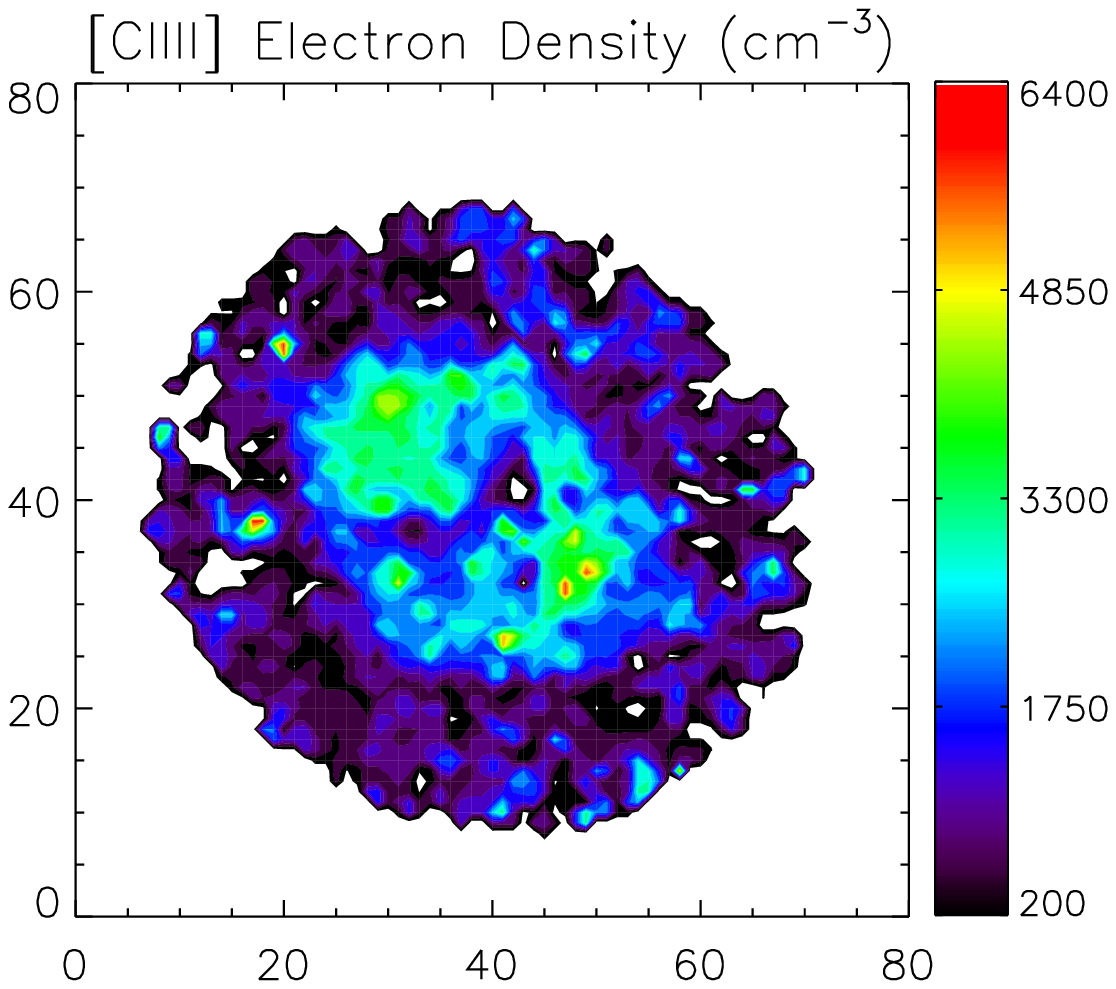}\\
\includegraphics[width=\columnwidth]{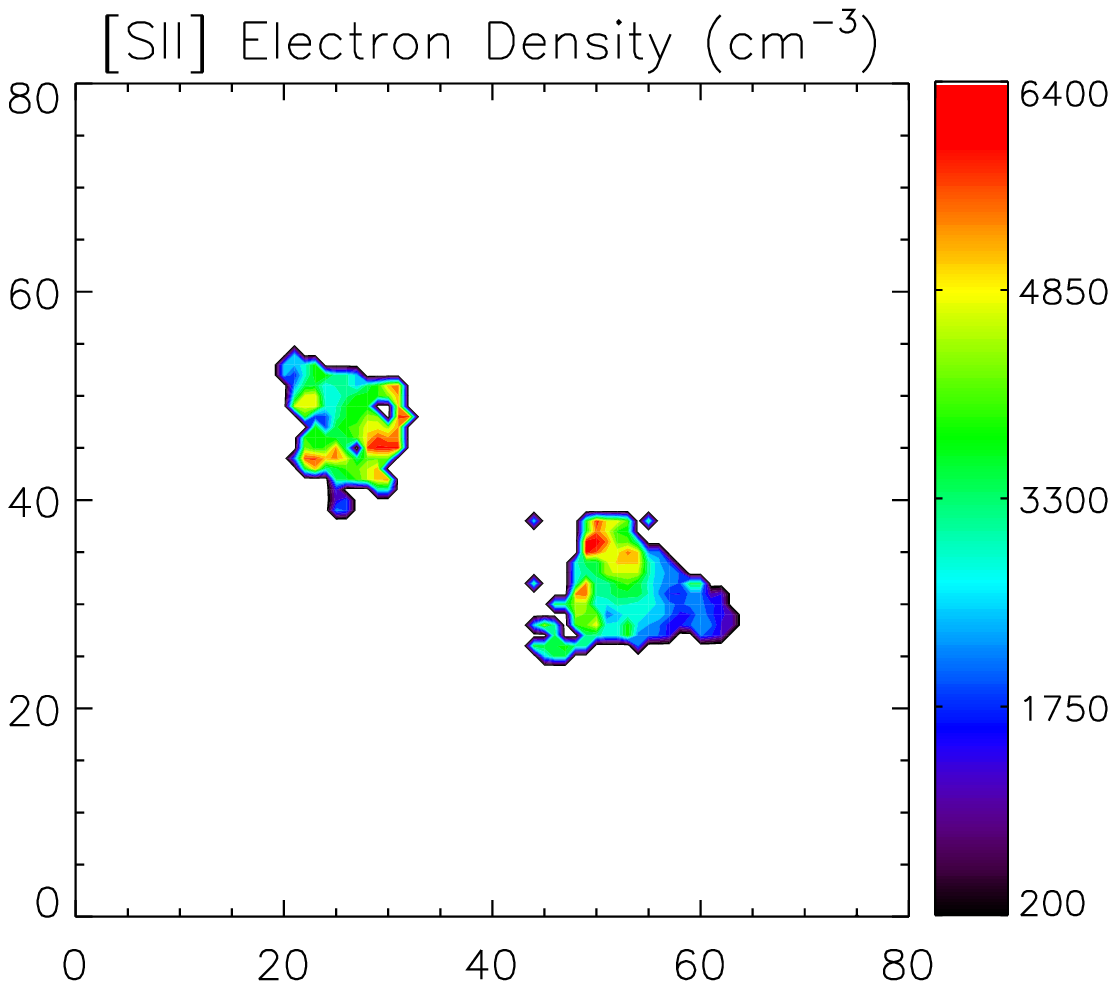}
\caption{Diagnostic ratio maps for the electron density derived from
  the [Ar~{\sc iv}], [Cl~{\sc iii}] and [S~{\sc ii}]
  emission-lines. The orientation in the sky as well as the
    plate scale are the same as in
    Fig. \ref{lna-slit}}\label{ne-maps}
\end{center}
\end{figure}

The extinction-corrected emission-line maps were used to derive the
spatial distribution of the electron density (N$_e$) and temperature
(T$_e$) of the nebula. By assuming a constant T$_e$ throughout the
nebula we obtained a first estimate for the N$_e$ maps based on the
[S~~{\sc ii}], [Cl~~{\sc iii}] and [Ar~~{\sc iv}] lines. The estimated
maps were used in an iterative procedure to obtain the final N$_e$ and
T$_e$ maps. The results for the density are shown in
Figure~\ref{ne-maps}. The mean values of the electron densities,
weighted by the most intense line map used in the ratio, were
N$_e$[Ar~{\sc iv}]=$2,640$~cm$^{-3}$, N$_e$[Cl~{\sc
  iii}]=$2,380$~cm$^{-3}$, and N$_e$[S~{\sc
  ii}]=$3,120$~cm$^{-3}$. Since most of the works in the literature
reported values obtained from integrated fluxes, we present these
quantities and their associated errors in Table~\ref{ne-tab}, together
with results taken from the literature.  Considering that errors are
not quoted in all previous works shown in Table~\ref{ne-tab}, our
means agree reasonably well with the other density estimations.

\begin{figure}[!ht]
\begin{center}
\includegraphics[width=\columnwidth]{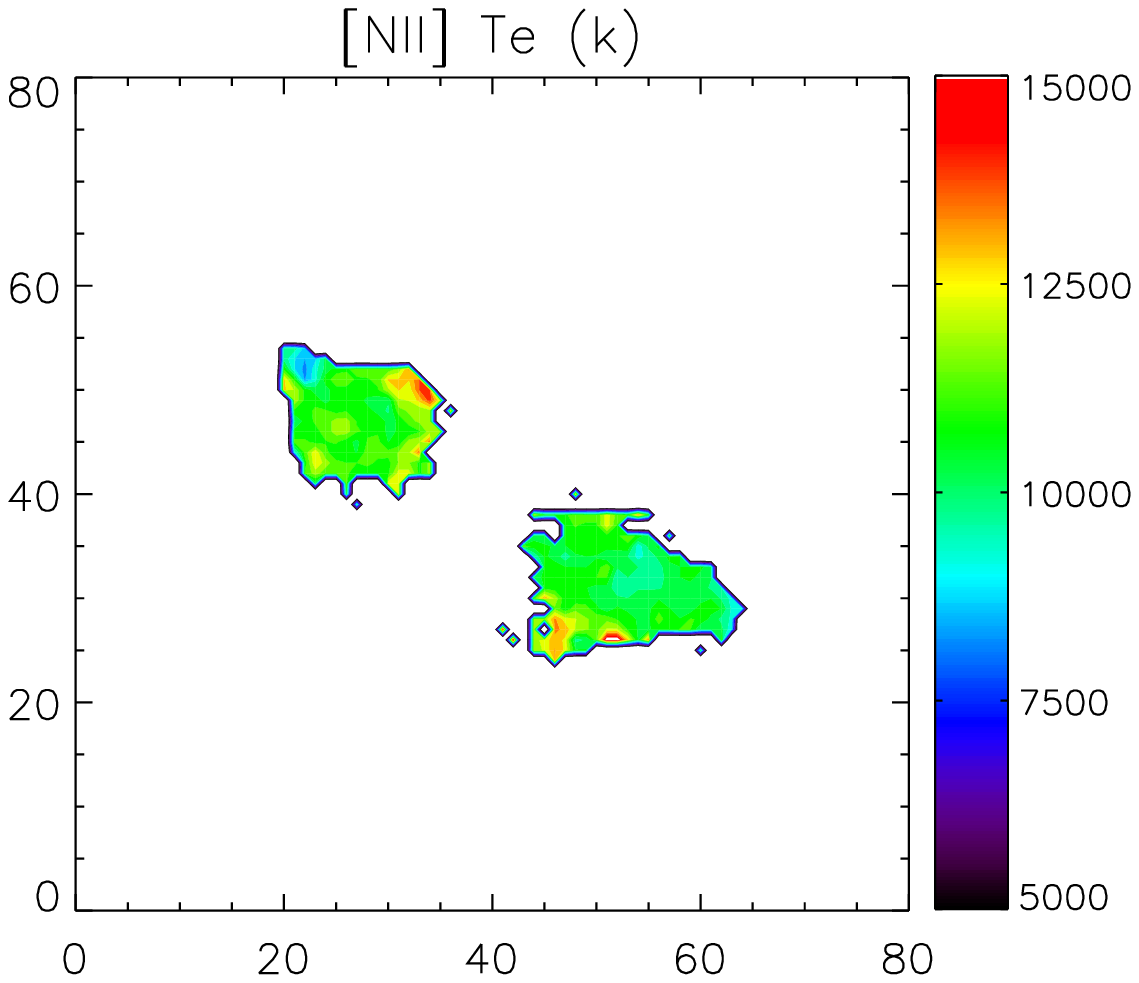}\\
\includegraphics[width=\columnwidth]{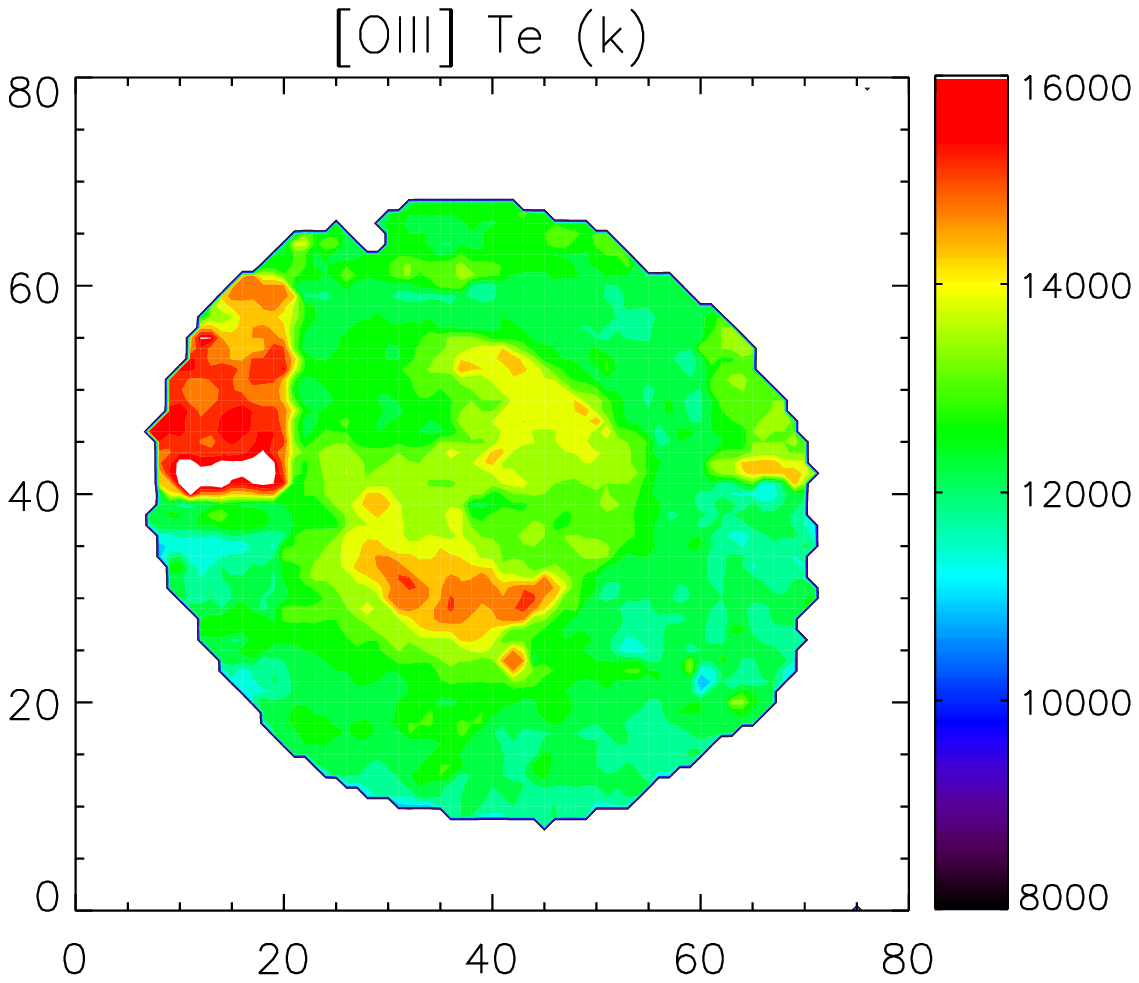}
\caption{Diagnostic ratio maps for the electron temperature derived
  from the [N~{\sc ii}] and [O~{\sc iii}] emission maps.  The
  orientation in the sky as well as the plate scale are the same as in
  Fig. \ref{lna-slit}}\label{te-maps}
\end{center}
\end{figure}

\begin{table}[!ht]
  \caption{Total line fluxes relative to $H\beta$ obtained from the emission-line maps and compared with the total fluxes of Tsamis et al. (2003; T03).}             
  \label{table:1}      
  \centering           
  \begin{tabular}{l c l l}
\hline        \\    
Emission line &   Wavelength (\AA) & this work &          T03 \\ \\
\hline \\                 
\ \ H$\gamma$      &4340 &    0.434    &   0.461 \\
\ \ [O~{\sc iii}]  &4363 &    0.130    &   0.132 \\
\ \ He~{\sc ii}    &4686 &    0.109   &   0.255 \\
\ \ [N~{\sc iii}]  &4640 &    0.047  &   0.016 \\
\ \ [Ar~{\sc iv}]  &4711 &    0.067   &   0.049 \\
\ \ [Ar~{\sc iv}]  &4740 &    0.047  &   0.045 \\
\ \ H$\beta$       &4861 &  1.000	 &   1.000 \\
\ \ [O~{\sc iii}]  &5007 &    12.87     &	13.00 \\
\ \ He~{\sc i}     &5876 &    0.18    &   0.122 \\
\ \ H$\alpha$      &6563 &  4.37	  &   3.100 \\
\ \ [Cl~{\sc iii}] &5517 &    0.0038      &   0.0033 \\
\ \ [Cl~{\sc iii}] &5537 &    0.0036      &   0.0029 \\
\ \ [N~{\sc ii}]   &5755 &    0.0007      &   0.0007 \\
\ \ He~{\sc i}     &6678 &    0.039   &   0.034 \\
\ \ [N~{\sc ii}]   &6548 &    0.011  &   0.011 \\
\ \ [N~{\sc ii}]   &6584 &    0.031   &   0.028 \\
\ \ [O~{\sc i}]    &6300 &    0.0003 &   0.0005 \\
\ \ [S~{\sc ii}]   &6717 &    0.0016  &   0.003 \\
\ \ [S~{\sc ii}]   &6731 &    0.0024  &   0.004 \\
\ \ [S~{\sc iii}]  &6312 &    0.008  &   0.007 \\
\hline                                   
\end{tabular}
\label{flux-tab}
\end{table}

\begin{table}[!ht]
  \begin{center}
    \caption{ Densities and temperatures derived from typical
      diagnostic line ratios, from our maps (mean values) compared to
      with some selected previous works (Tsamis et al 2003, T03;
      Krabbe \& Copetti 2006, KC06; Pottasch \& Bernard-Salas 2008,
      PB08).}
\begin{tabular}{lllll}
\hline \\
 Diagnostic                     & this work     &  T03  & KC06  & PB08   \\ 
 (cm$^{-3}$/K) & mean& && \\
 \\
\hline \\
N$_e$[Ar~{\sc iv}]   & $2,640 \pm 400$     &	 3,040 & $3,665 \pm 141$  & 2,100  \\
N$_e$[Cl~{\sc iii}]  & $2,380 \pm 1,300$   &   1,200 & $2,531 \pm 1,227$ & 3,000  \\
N$_e$[S~{\sc ii}]    & $3,120 \pm 1,400$   &   1,970 & $1,016 \pm 436$  & 1,900  \\
T$_e$[O~{\sc iii}]          & $12,900 \pm 720$     &	11,700 & $12,140 \pm 31$  & 10,800 \\
T$_e$[N~{\sc ii}]            & $11,670 \pm 1000$     &	13,400 & $--$             & 11,000 \\
\hline 
\end{tabular}
\label{ne-tab}
\end{center}
\end{table}

To derive the electron temperature maps, we used the N$_e$ maps
previously described (Fig.~\ref{ne-maps}) as input parameters. To
obtain the final temperature maps the calculations were performed
pixel by pixel, each with its respective density value. As usual, the
similarity of the ionisation potential was the criterion for adopting
T$_e$[N~{\sc ii}] or T$_e$[O~{\sc iii}], and [S~{\sc ii}] or [Ar~{\sc
  iv}] densities.  Our temperature maps are presented in
Fig.~\ref{te-maps} and correspond to mean values of 11,670~K and
12,900~K for T$_e$[N~{\sc ii}] and T$_e$[O~{\sc iii}],
respectively. The figures show that the temperature distribution is
very constant in both maps apart from an increase in the rim region
seen in the T$_e$[O~{\sc iii}] map. Within the rim, however, the
temperature shows little fluctuation. The temperatures found in our
maps are consistent with values determined by \cite{monreal05}
within their quoted uncertainties.

\subsection{Ionic and total chemical abundances}

The N$_e$ and T$_e$ maps were used as input to determine ionic
abundances.  They are accounted for such that low- (T$_e$[N~{\sc ii}],
N$_e$[S~{\sc ii}]) and medium- (T$_e$[O~{\sc iii}], N$_e$[Ar~{\sc
  iv}], N$_e$[Cl~{\sc iii}]) excitation electron temperatures and
densities are, correspondingly, adopted for the low- ([O~{\sc i}],
[S~{\sc ii}] and [N~{\sc ii}]) and intermediate-ionisation ions
([O~{\sc iii}], [Cl~{\sc iii}], [S~{\sc iii}] and [Ar~{\sc iv}]). The
ionic abundance obtained in this work from the mean fluxes of
Table~\ref{ne-tab} values are show in Table~\ref{tab-abund}, where
those from the literature are also given for comparison.

To obtain the total chemical abundance maps we followed the
prescriptions by Kingsburgh \& Barlow (1994) when only optical data is
available. These authors provided the ionisation correction factors
(ICF) that account for the ionic maps we cannot calculate from our
data.  For oxygen, a key ionic map, O$^+$/H$^+$, is missing, because
the VIMOS configuration does not include either the [O~{\sc ii}]
3727\AA\ or the 7325\AA\ emission lines. Therefore, we generated ionic
abundance maps from values found in the literature
(2.54$\times$10$^{-6}$ and 4.16$\times$10${^-6}$ for O$^+$/H$^+$ based
on 3727\AA\ and 7325\AA, respectively, T03) and assumed that these
abundances are uniform throughout the map.  We then calculated the
average of these two uniform maps (=3.35$\times$10$^{-6}$), generating
a final O$^+$/H$^+$ that we used to obtain the ICFs. This was
necessary because without this ionic map, the total abundance of
oxygen (O/H) cannot be calculated, and as a consequence,the entire ICF
would be missing (see Kingsburgh \& Barlow 1994, Appendix A).

Maps showing the total abundances of oxygen, nitrogen, sulphur,
chlorine, and helium are shown in Fig.~6.  By examining Fig.~6 we see
no significant abundance variation of O, N, S or Cl (based on
uncertainty estimates obtained in Sec. 4). It is also worth noting
that for some elements only one ionic fraction is available. In these
cases, the total abundances were determined only for the regions for
which the corresponding emission-line is strong. The extreme example
of this effect is observed for N$^+$/H$^+$. In
Fig.~6 
the strong LISs are the only structures for which we were able to
derive the N/H.

\begin{figure*}[!ht]
\begin{center}
\includegraphics[width=\columnwidth]{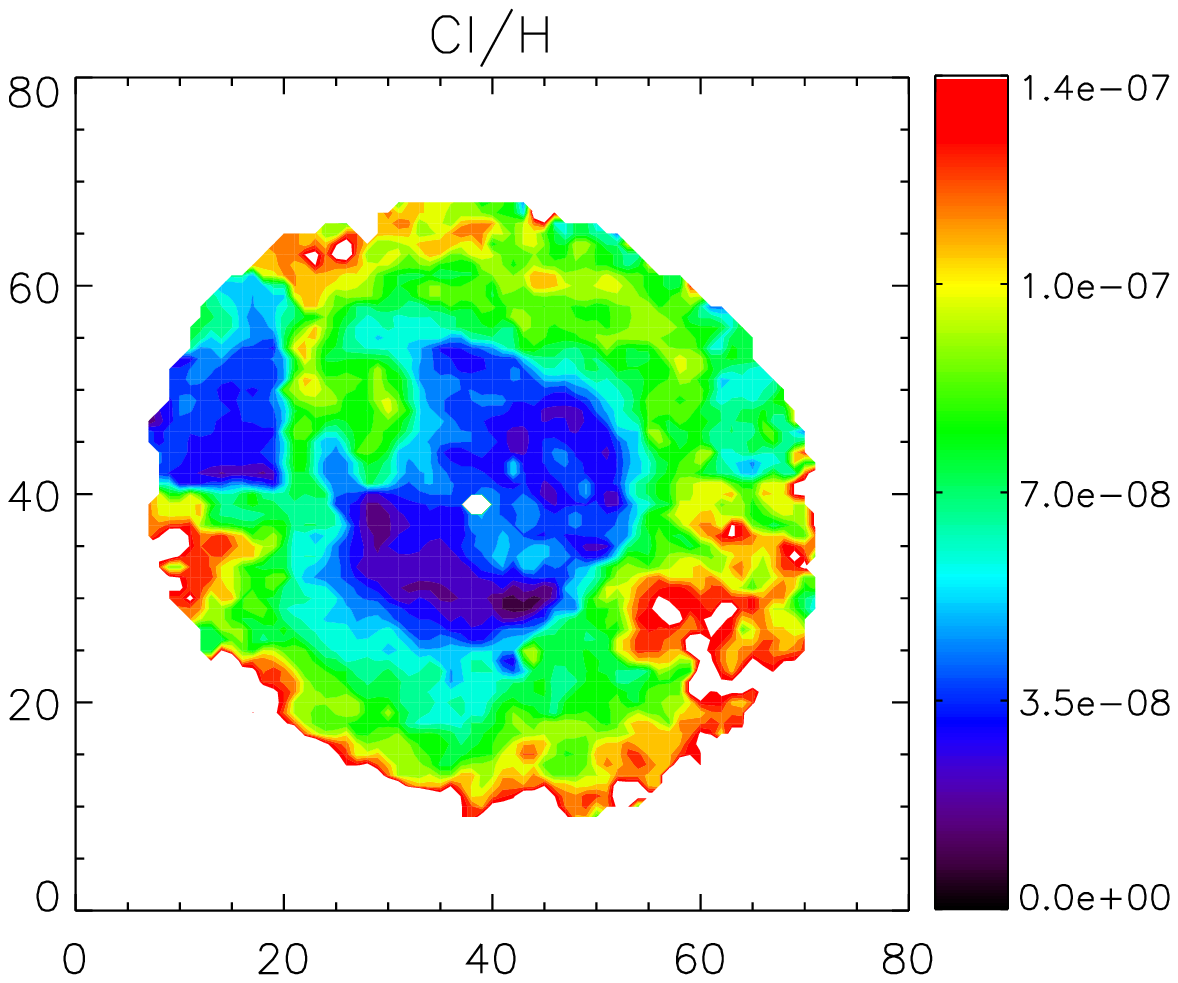}
\includegraphics[width=\columnwidth]{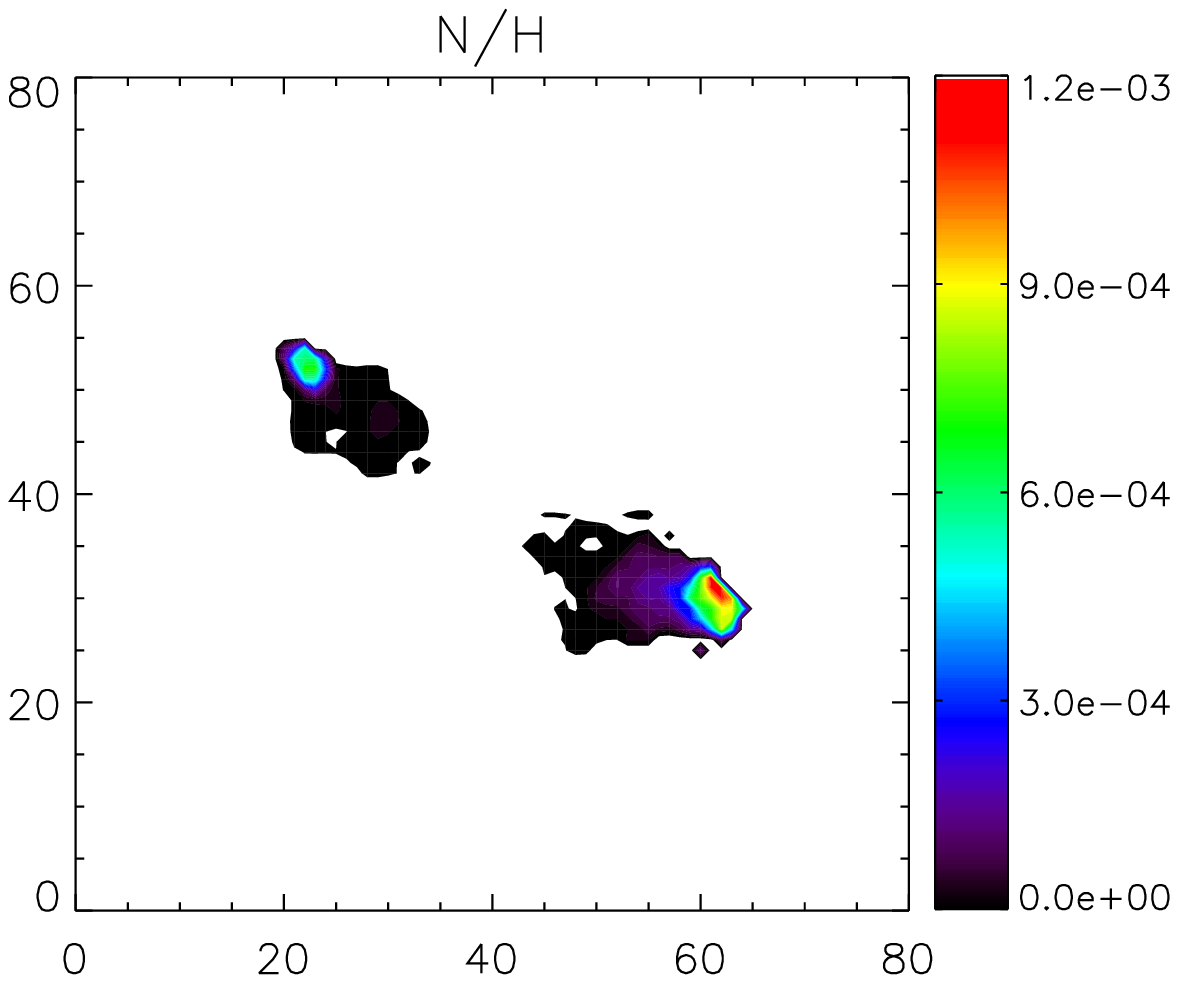}\\
\includegraphics[width=\columnwidth]{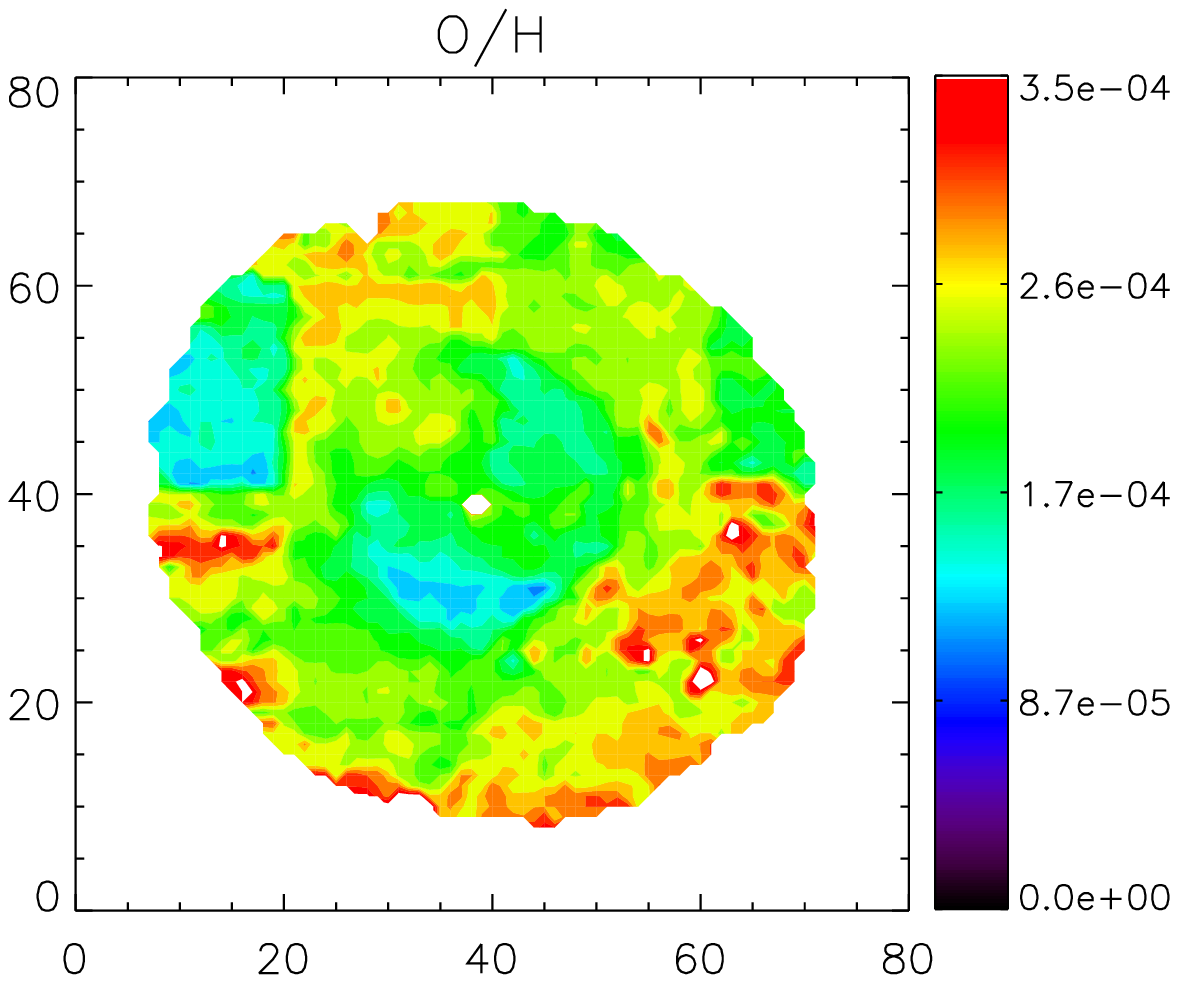}
\includegraphics[width=\columnwidth]{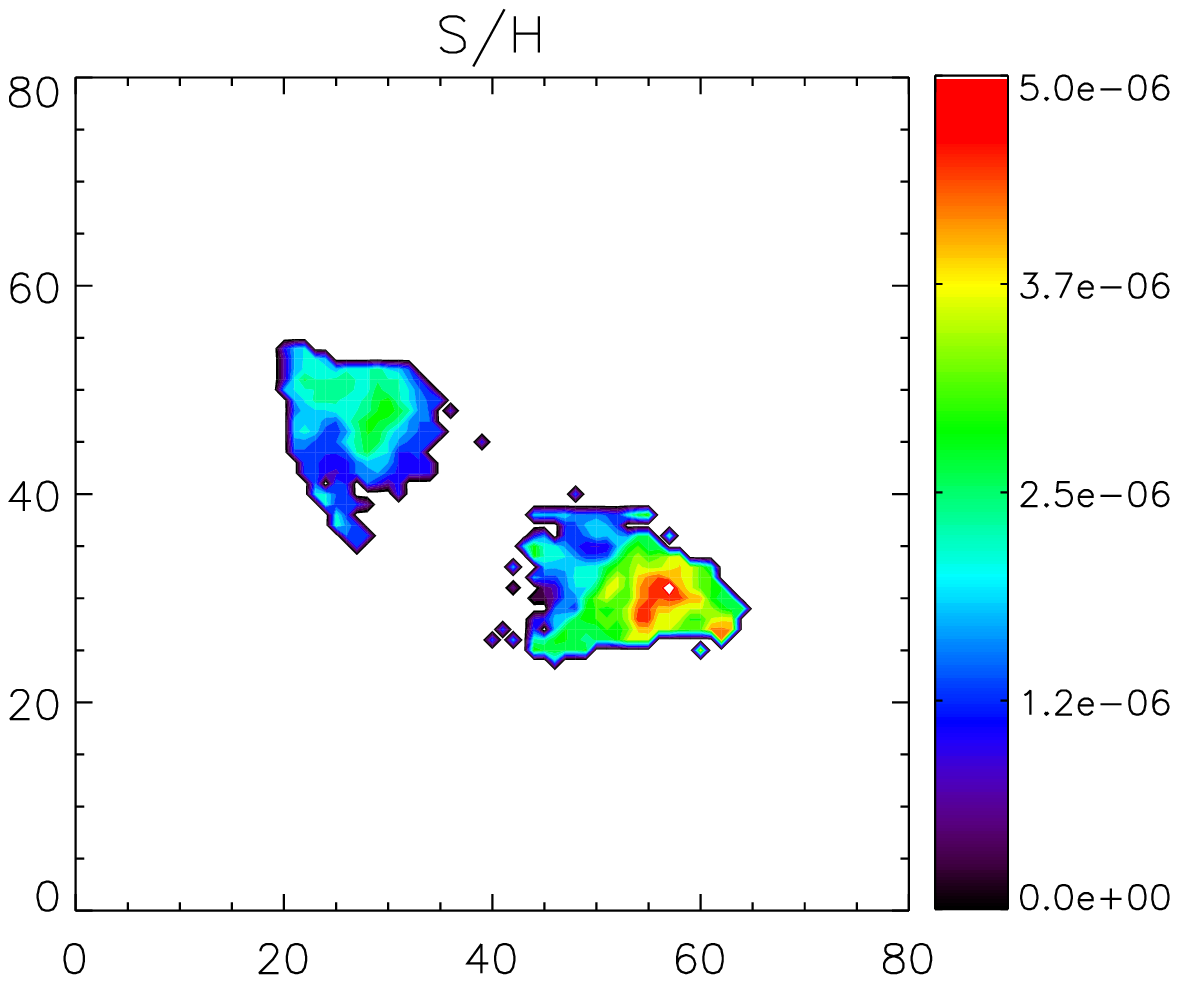}\\
\includegraphics[width=\columnwidth]{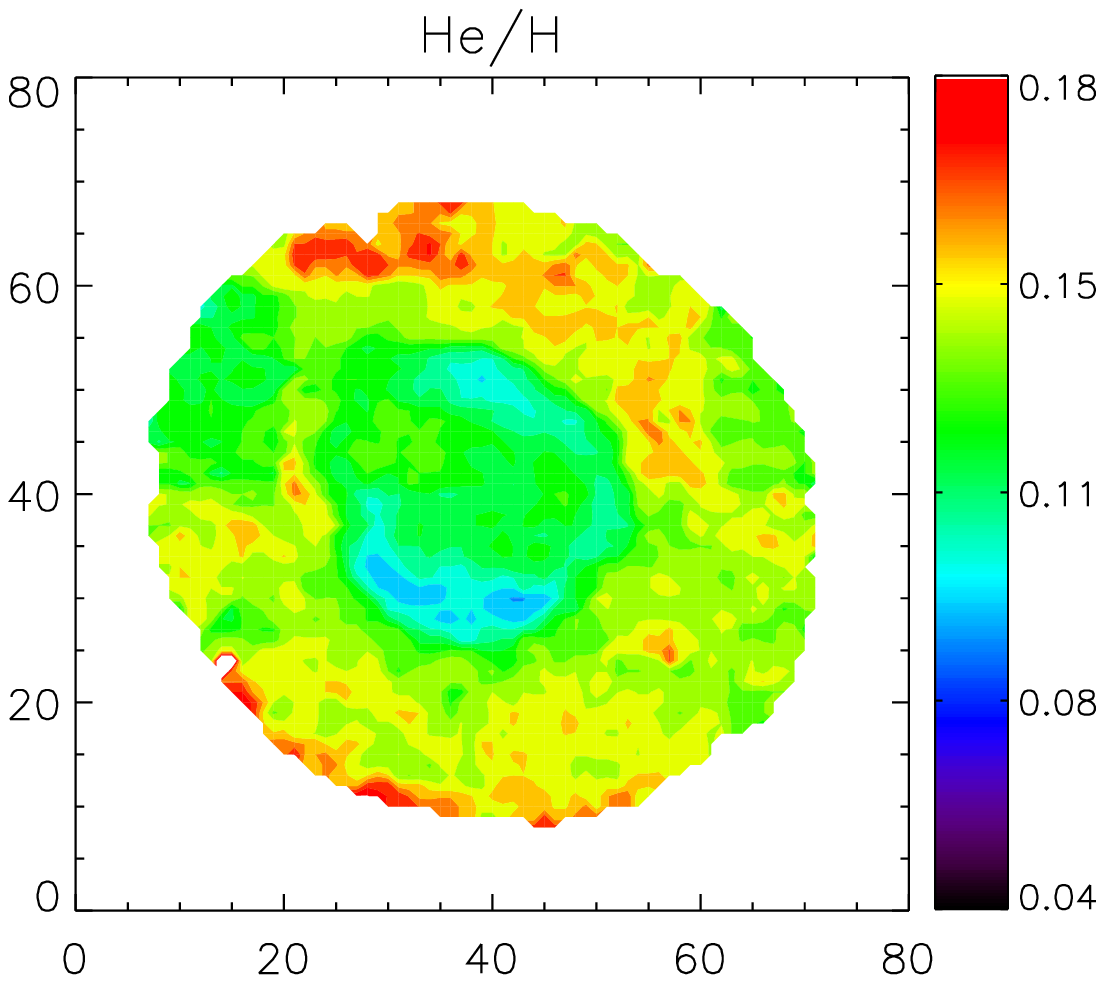}

\caption{Total abundance maps. The orientation in the sky
    as well as the plate scale are the same as in
    Fig. \ref{lna-slit}}
\end{center}
\label{abund-maps}
\end{figure*}

Several authors have obtained ionic and total abundances of He and the
most abundant heavy elements of NGC~3242, from long-slit
spectra. Three of these abundances results are shown in
Table~\ref{tab-abund} and are compared with our own calculated values,
estimated from the total line fluxes.  We point out that the
wavelength coverage as well as the region covered by the slit/aperture
vary in the literature. Krabbe \& Copetti (2006) data covering the
wavelength range from 3,100\AA\ to 6,900\AA, is closer to ours not
only in terms of spectral coverage, but also in terms of the elemental
abundances and ICFs derived although they use a single slit. Similar
to our studie, their study was based on optical spectra alone. At the
other extreme of reported values are those of Pottasch \&
Bernard-Salas~(2008), who used the Spitzer Space Telescope and the
Infrared Space Observatory for the near-IR wavelengths and the
International Ultraviolet Explorer (IUE) as well as ground-based
optical spectra, all of which have distinct apertures and slit
sizes. It is worth noting that they needed no ICFs in their analysis,
therefore their elemental abundances should be the best. The
comparison shows that the Pottasch \& Bernard-Salas~(2008) results
differ from ours by more than the other two sets of results. But on
the whole, the difference in the abundances obtained in all the four
papers is always small. The exception is the N abundance, an element
for which the optical abundance is well-known to be poorly
determined. The only way of improving the N/H abundances is by adding
the UV spectrum to the optical and/or optical- plus- near-IR
spectrum. Tsamis et al. (2003) scanned the slit across the nebula in
the optical spectrum acquisition and added to the analysis IUE
data. It is straightforward to see from Table~\ref{tab-abund} that
their results are similar to ours, again with the exception of the
nitrogen abundance.

\begin{table*}[!ht]
  \begin{center}
    \caption{Ionic and total abundances obtained from integrated line
      fluxes compared with those obtained in the literature.}

\begin{tabular}{llllll}\hline \\
                & This work$^a$ &        &Tsamis et al. (2003)$^a$ & Krabbe \& Copetti (2006)$^b$ & Pottasch \& Bernard-Salas (2008)$^c$\\ \\ \hline \\
Ion             & ICF & Abundance    & Abundance (ICF)   & Abundance (ICF)    & Abundance (ICF) \\ \\ \hline \\
He$^+$/H    &    & 0.08  &                   &	              & \\ 
He$^{++}$/H  &    & $9.45\times 10^{ -3}$ &                   &	              &\\ 
{He/H}       &    & 0.09	     & 0.10               & 0.10               & 0.092\\ \\

O$^0$/H    &    & $2.11\times 10^{ -8}$ &                   &	              &\\
O$^{2+}$/H  &    & $2.34\times 10^{ -4}$ &                   & 	              &\\
{\b O/H }       &1.07& $2.55\times 10^{ -4}$ & $3.3 \times 10^{-4}$~(1.17)& $3.09\times 10^{-4}$~(1.30)& $3.80 \times 10^{-4}$~(1.00)\\  \\

N$^{+}$/H   &    & $2.63\times 10^{ -7}$ &                   &     	              &\\
{N/H }       &80.00& $2.05\times 10^{ -5}$ & $3.4 \times 10^{-5}$~(1.53)& $4.19 \times 10^{-5}$~(220.)& $1.35 \times 10^{-4}$~(1.00)\\  \\

S$^{+}$/H     &    & $6.02\times 10^{ -9}$ &                   &	              &\\
S$^{2+}$/H    &    & $7.43\times 10^{ -7}$ &                   &	              &\\ 
{S/H }       &3.00& $2.24\times 10^{ -6}$ & $2.4 \times 10^{-6}$~(3.52)& $2.60 \times 10^{-6}$~(4.20)& $2.80 \times 10^{-6}$~(1.00)\\   \\                  

\hline
\end{tabular}
\label{tab-abund}
\end{center}
\tiny
 $^a$ covered full nebula;\\
$^b$ used single long-slit with 2'' ;\\
$^c$ used data from different sources with different aperture or slit sizes.
\end{table*}

Though within the range of expected values, the He/H abundance
variations in the map are significant. To investigate this in more
detail, we compared our mean He/H with the other values for the entire
nebula in the literature (see Table~\ref{tab-abund}). Inspecting the
table, we see that our He/H, agrees with the other values reported in
the literature , which indicates that our results are good at least on
average.

Although they agree, within the errors, the total line fluxes
and the differences between our values and those obtained by T03 may
indicate that the current precision and accuracy obtained with the
typical observing techniques may not be adequate for the investigation
of spatial variation of chemical abundance.

\section{Zone Analysis: summing up the emission per region}

\begin{figure}[!ht]
\begin{center}
\includegraphics[width=\columnwidth]{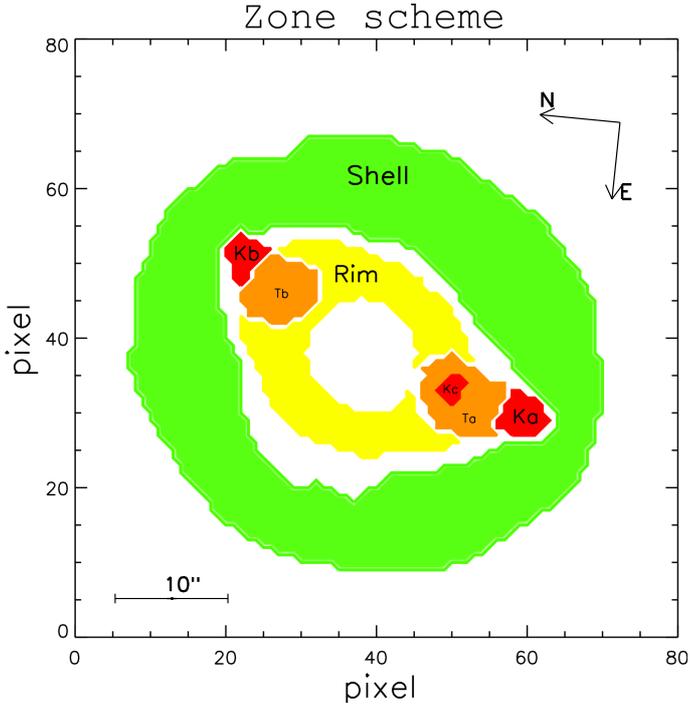}
\caption{Zone scheme definitions for NGC 3242. The orientation in
    the sky as well as the plate scale are the same as in
    Fig. \ref{lna-slit}}\label{zones}
\end{center}
\end{figure}

As discussed above, a study of spatial abundance variations using the
maps obtained, that are displayed in Fig. \ref{abund-maps} is made
difficult by the limited S/N ratio in individual pixels.  However, IFU
data have the advantage that the signal can be integrated over given
sets of pixels, for instance covering specific morphological
components of the nebula.  Given the symmetry of the object, we used
isophotal contours to delineate the boundaries of nebular regions of
particular interests.  Specifically, we defined seven distinct zones
that separate all of the major morphological features seen in the
emission line maps. The zones and their boundaries are shown in the
schematic Fig.~\ref{zones}. The largest and faintest zone is refered
to as the shell. The second zone, commonly refered to as the
rim, is the brightest and densest structure of the nebula.

\begin{figure*}[!ht]
\begin{center}
\includegraphics[scale=0.65]{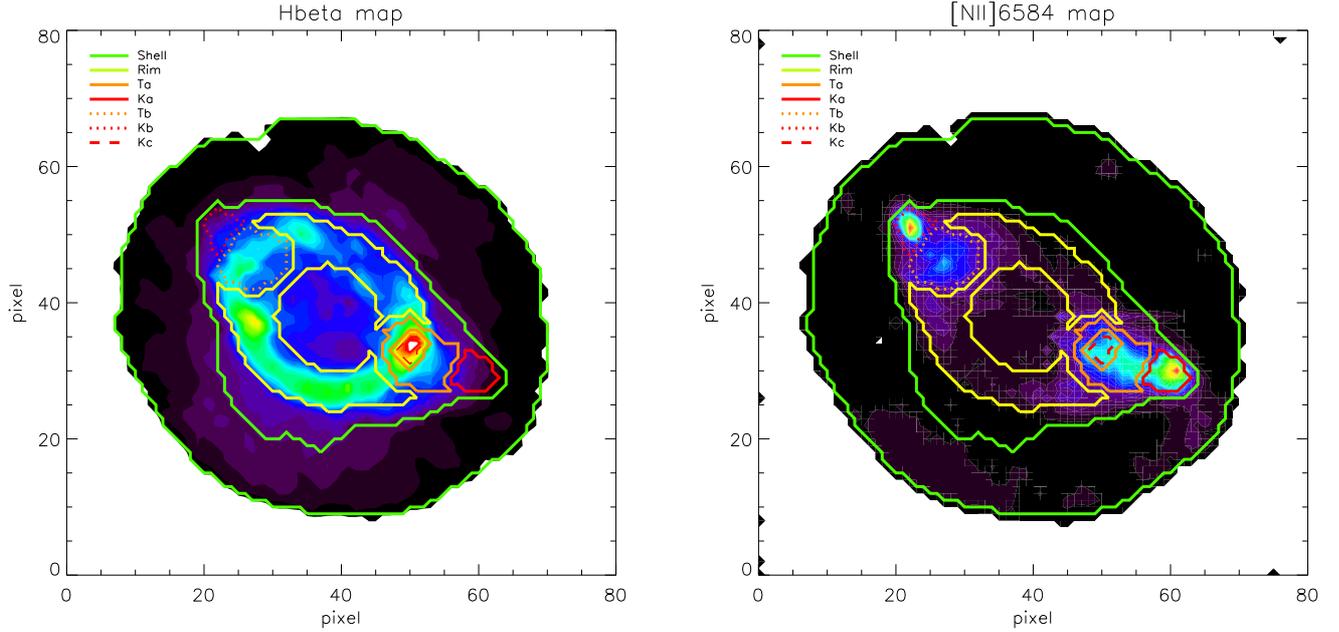}
\caption{Zone contours overlaid on observed emission-line maps of
  $H\beta$ (left) and $[N~{\sc II}]6584$ (right). The orientation in
  the sky as well as the plate scale are the same as in
  Fig. \ref{lna-slit}}\label{zones-hbn2}
\end{center}
\end{figure*}

The other zones were chosen to attend to the region of the bright
knots (LISs) seen in images. Here we opted to divide the knot into two
main parts in an attempt to isolate the knot from the other structures
as much as possible. These two regions, defined for the two visible
knots, are composed of the knots themselves and a tail that
morphologically seems to lag behind the actual knots. In this way, the
knots in principle contain the lowest possible contamination from the
other zones and the tail also does not contain most of the emission of
the knots. The tail zones overlap with the rim to some extent. The
tails are are better visible in Fig. \ref{zones-hbn2}, where we show
the zone contours overlaid on the emission maps of H$\beta$ and
[N~{\sc ii}]6584\AA. The tail and knot zones are defined such that the
south knot is connected to zones tail a and knot a and the north knot
with zones tail b and knot b. We also defined a final zone {knot c}
connected to a structure that is very bright, in most emission-lines,
and is situated at the bottom of tail a and on top of the rim.

For each of these zones, we sumed up the fluxes and performed the
same analysis as in the previous section. In Table~\ref{table:1} we present 
the total extinction-corrected fluxes, densities, temperatures, ionic
abundances, and total abundances for each zone.

\begin{table*}[!ht]

  \caption{Physical parameters, ionic and total abundances, and ICFs obtained 
    for each zone defined from isophotal contours.}             

  \label{table:1}      
  \centering           
  \begin{tabular}{lccccccc}
\hline             
      \\
Line ID & shell & rim & { tail a (Ta)} & { knot a (Ka)} & { tail b (Tb)} & { knot b (Kb)} & { knot c (Kc)} \\
      \\
\hline                  


\hfill \\
 Density diagnostic ($cm^{-3}$)& & & & & & &  \\
 
\ \  [Ar~{\sc iv}]   &	352. &   2250. &  1450. &  720. &   1170. &  400. &  2360. \\
\ \  [Cl~{\sc iii}]  &	768. &   1980. &  2120. &  1580. &  2750. &  1500. &  2500. \\
\ \  [S~{\sc ii}]    & 1794. &	 3920. &  3660. &  2190. &  5200. &  3450. &  5190. \\

\hfill \\
 Temperature diagnostic (K) & & & & & & &  \\
\ \  [N~{\sc ii}]   &  12639. &  14360. &  10120. &  9760. &  10500. &  8960. &  11130. \\
\ \  [O~{\sc iii}]  &  11673. &	 12430. &  11320. &  10970. &  12180. &  12070. &  11920. \\

\hfill \\
Ionic abundance & & & & & & &  \\

\ \  N$^{+}$/H	 &  $2.51\times 10^{ -7}$  & $1.08\times 10^{ -7}$ & $1.05\times 10^{ -6}$ & $6.25\times 10^{ -6}$ & $6.82\times 10^{ -7}$ & $3.84\times 10^{ -6}$ & $4.02\times 10^{ -7}$ \\
\ \  O$^0$/H	 &  $4.17\times 10^{ -8}$  & $5.22\times 10^{ -9}$ & $3.96\times 10^{ -8}$ & $3.33\times 10^{ -7}$ & --                    & $1.41\times 10^{ -7}$ & $1.13\times 10^{ -8}$ \\
\ \  O$^{+}$/H	 &  $3.08\times 10^{ -6}$  & $2.54\times 10^{ -6}$ & $8.95\times 10^{ -6}$ & $9.06\times 10^{ -6}$ & $8.98\times 10^{ -6}$ & $1.62\times 10^{ -5}$ & $7.29\times 10^{ -6}$ \\
\ \  O$^{2+}$/H	 &  $2.83\times 10^{ -4}$  & $2.03\times 10^{ -4}$ & $2.99\times 10^{ -4}$ & $3.66\times 10^{ -4}$ & $2.38\times 10^{ -4}$ & $2.56\times 10^{ -4}$ & $2.44\times 10^{ -4}$ \\
\ \  Ar$^{3+}$/H &  $9.21\times 10^{ -7}$  & $1.11\times 10^{ -6}$ & $1.04\times 10^{ -6}$ & $8.66\times 10^{ -7}$ & $8.70\times 10^{ -7}$ & $7.82\times 10^{ -7}$ & $9.78\times 10^{ -7}$ \\
\ \  S$^{+}$/H	 &  $6.65\times 10^{ -9}$  & $2.88\times 10^{ -9}$ & $2.17\times 10^{ -8}$ & $7.02\times 10^{ -8}$ & $1.66\times 10^{ -8}$ & $5.06\times 10^{ -8}$ & $1.00\times 10^{ -8}$ \\
\ \  S$^{2+}$/H	 &  $8.89\times 10^{ -7}$  & $6.33\times 10^{ -7}$ & $1.13\times 10^{ -6}$ & $1.41\times 10^{ -6}$ & $9.30\times 10^{ -7}$ & $8.49\times 10^{ -7}$ & $6.97\times 10^{ -7}$ \\
\ \  Cl$^{2+}$/H &  $3.50\times 10^{ -8}$  & $1.87\times 10^{ -8}$ & $3.44\times 10^{ -8}$ & $5.04\times 10^{ -8}$ & $2.78\times 10^{ -8}$ & $2.95\times 10^{ -8}$ & $2.05\times 10^{ -8}$ \\
\ \  He$^+$/H	&  0.13                  & 0.08                 & 0.12                 & 0.14                & 0.10                 & 0.11                 & 0.09 \\
\ \  He$^{++}$/H &  $9.26\times 10^{ -4}$  & 0.02                  & $6.83\times 10^{ -3}$ & $1.61\times 10^{ -4}$ & 0.01                  & $4.26\times 10^{ -3}$ & 0.01 \\

\hfill \\
ICF & & & & & & &  \\

\ \  N          &    93.06  &   92.71  &  35.78  &  41.52  &  29.46  &  17.19  &   37.33 \\
\ \  O          &    1.00   &   1.14   &  1.04   &  1.00   &   1.07  &  1.03  &   1.08 \\
\ \  S          &    3.15   &   3.15   &  2.31   &  2.42   &   2.17  &  1.82  &   2.34 \\
\ \  Cl          &       &      &     &     &    &  1.93  &  2.37  \\

\hfill \\
Total abundance & & & & & & &  \\
 
\ \  He/H       &  0.13                 & 0.10                &  0.13      & 0.14  &  0.11  &  0.12 & 0.10 \\
\ \  N/H        &  $2.34\times 10^{ -5}$ & $1.00\times 10^{ -5}$ & $3.77\times 10^{ -5}$ & $2.59\times 10^{ -4}$ & $2.01\times 10^{ -5}$ & $6.60\times 10^{ -5}$ & $1.50\times 10^{ -5}$ \\
\ \  O/N        &  $2.87\times 10^{ -4}$ & $2.35\times 10^{ -4}$ & $3.20\times 10^{ -4}$ & $3.76\times 10^{ -4}$ & $2.64\times 10^{ -4}$ & $2.79\times 10^{ -4}$ & $2.72\times 10^{ -4}$ \\
\ \  S/H        &  $2.82\times 10^{ -6}$ & $2.00\times 10^{ -6}$ & $2.66\times 10^{ -6}$ & $3.59\times 10^{ -6}$ & $2.05\times 10^{ -6}$ & $1.64\times 10^{ -6}$ & $1.65\times 10^{ -6}$ \\
\ \  Cl/H       &  $1.11\times 10^{ -7}$ & $5.91\times 10^{ -8}$ & $8.10\times 10^{ -8}$ & $1.28\times 10^{ -7}$ & $6.13\times 10^{ -8}$ & $5.71\times 10^{ -8}$ & $4.88\times 10^{ -8}$ \\

\hline                                   
\end{tabular}
\end{table*}

Some of the previous works on NGC~3242 discussed above, which have
data for the entire nebula (c(H$\beta$), T$_e$, N$_e$ and abundances),
also reported spatially resolved estimates of physical properties for
the rim, shell and for the {knots}. The most recent of
these works is that of \cite{ruiz11} who pesented temperatures of
10,000 $\le$ T$_e$(K) $\le$ 14,700 (8,070 $\le$ T$_e$(K) $\le$ 10,400)
for the rim (shell). Although formal errors are not
quoted, the general picture from the comparison of our results to this
one is that their electron temperatures are systematically lower than
ours. The discrepancy for the temperature is 1,200~K and 2,000~K when
derived from the [N~{\sc ii}] and [O~{\sc iii}] ratios, respectively.
We note that these differences are smaller than the dispersion in
temperature due to the different emission-line ratios, which means
that they are probably constrained within the errors. These authors
derived, for the rim and shell, values for the electron
densities that vary from 2,200 to 2,250~cm$^{-3}$ and 340 to
400~cm$^{-3}$, respectively. The two density diagnostics they used
were [S~{\sc ii}] and [Ar~{\sc iv}], while we used [Cl~{\sc iii}] as
well. In general, despite the different observational data and zone
definitions, the results agree within the possible uncertainties
which, for faint lines such as the [S~{\sc ii}], can reach 50\%. The
main differences are in the densities obtained from the [S~{\sc ii}]
lines for the shell, but in this case our values should be taken
with care, because of very low S/N for the line in this particular
zone. The effect can be clearly seen in the [S~{\sc ii}] density map
shown in Fig.4. In the areas with a good S/N for the line the density
values agree with those of the other authors.

Another comparison can be made with the long-slit data reported
\cite{balick93}. In that work, authors constructed a spatially
resolved study of NGC 3242 and specifically discussed the {
  knot}. They derived [N~{\sc ii}] and [O~{\sc iii}] electron
temperatures, and [S~{\sc ii}] and [Cl~{\sc iii}] densities for the
four components of the nebula for which we derived the nebular
properties shown in Table~4. Their Fig. 4 shows profiles of these
quantities evaluated along the long-slit
(P.A.=~-30$^\circ$). According to this figure, T$_e$[O~{\sc iii}]
(T$_e$[N~{\sc ii}]) of both { knots} are 10,000 to 11,000~K
($\sim$9,000~K). These figures agree well with our results, especially
for T$_e$[N~{\sc ii}]. The trend of higher T$_e$[O~{\sc iii}] than
T$_e$[N~{\sc ii}] is found in both works as well. \cite{balick93}
found T$_e$[O~{\sc iii}]=11,000~K at the rim and shell and
quoted that their uncertainty is of about 10\%, which makes their
results similar to ours apart from a higher T$_e$[N~{\sc ii}] at the
rim. However, this particular value should be taken with care
because the highly localised emission from the [N~{\sc ii}]
lines. They interpreted their results as indicating constant electron
temperatures throughout the nebula. When densities are the focus, the
values we found for the rim and shell are about half of
those reported in \cite{balick93} for the two emission-line
diagnostics in common. Despite the discrepancy, the values are within
the large uncertainty mentioned above. The larger discrepancy is
perhaps for { knot~b} where we find a density of 3,450~cm$^{-3}$,
which is similar to the rim density, again in contrast to the
trend that was found by \cite{balick93}. For this particular knot, due
to its proximity to the rim, it is likely that some
contamination from the nearby emission is present.

Indeed, the density map in Fig. 3 shows that the inconsistency
disappears in the more precise pixel-by-pixel analysis for regions
with a good S/N.  The results indicate that although summing pixels in
pre-defined zones may improve the S/N, it does not deal completely
with the contamination from nearby zones. It is very likely that the
{ knot~a} values for physical and chemical composition are more
representative of the knots, because this structure it is less
contaminated by the emission from the bright rim zone, as
indicated by the densities found.

The results of our analysis for the abundance variations within
NGC~3242, are also shown in Table~4. We found no significant abundance
contrast for He and O. The extreme case is N/H, whose shell
abundance is more than twice that of the rim, and the {
  knots} are up to six times more abundant than the rim. The
derived increase in abundance is apparent in both knots, indicating
that the result is not influenced by nearby emission
contamination. However, as discussed in Section~3.2, for this element
the ICFs are typically very high (varying from 93 in the rim to
42 in { knot~a}), which yields a very poor determination of the
abundances because most of the nitrogen ionic fractions are beyond the
optical spectrum (also see \cite{gon06}; \cite{gon12}), indicating
that the increases found are not reliable.

To obtain an estimate of the uncertainties involved in the abundance
determinations we performed a Monte Carlo analysis using the total
fluxes in Table 1. We calculated the abundances 5000 times, each time
sampling a different flux value from a Gaussian distribution with the
line flux as the mean and 5\% and 10\% total uncertainties (flux
calibration+photon statistics) for strong and weak lines,
respectively. The abundance $1\sigma$ uncertainties obtained are
$\approx 10\%$ for He, $\approx 15\%$ for O, $\approx 35\%$ for S, and
$\approx 60\%$ for N. Clearly, since we used the total fluxes, these
error estimates are lower limits and are probably higher for regions
of low S/N levels. When evaluating the results, however, it is
important to point out that the adequacy of the ICFs used cannot be
easily incorporated into the final uncertainties obtained, which
essentially means that the actual errors can be even higher.

Finally, taking into account all the discussions and caveats pointed
out above, and conservatively analysing Table~4, we find no
abundance variations for the elements He and O in the rim, {
  shell} and {knots} of NGC~3242. We also infer that the abundance
variation found for N ($>$6$\times$), S ($>$2$\times$), and Cl
($>$2$\times$) are probably caused by the high ICFs, which prevents us
from stating weather these results reflect the real conditions within the
nebula.

\section{Discussion and conclusions}

We investigated the spatially resolved physical and chemical
properties of the planetary nebula NGC~3242. The determination of the
physical and chemical properties of the object obtained from our
resolved IFU observations are in general consistent with those
obtained in the literature from data with more limited spatial
coverage.  The comparison of the total fluxes obtained in this work
with those of Tsamis et al. (2003) showed that our results agree well
despite the indirect flux calibration we had to adopt, thus validating
the procedure. The spatially resolved maps for the physical and
chemical conditions show considerable structure. When considering the
abundance maps, however, we did not infer significant abundance
variations within the limits of the present data.

To investigate the abundance variations from another angle using the
same data set, we also divided the nebula into zones for which we
obtained total fluxes. The zones isolated the main morphological
features of NGC~3242: the shell, the rim, and two {knots}. The results for the zone analysis corroborates other results
from the literature and shoed no evidence for significant spatial
variation of the chemical abundances of He and O, even in peculiar
features such as the {knots}. In these features, we found an
overabundance of N and S, but the uncertainties as well as the large
ICFs used showed that the increase is not significant. Ideally, proper
3D modelling is needed to analyse the highly asymmetric structures as
was performed in \cite{gon06}).

Several observational studies have investigated the question of
possible spatial variations of the abundances in PNe using long-slit
optical data. An example is the study of \cite{pc98}, who analysed
thirteen PNe with a bipolar morphology. These authors found that the
He, O, and N abundances were constant throughout all the nebulae
studied. The abundances of Ne, Ar, and S were also found to be
constant within the errors, but their face values implied a systematic
increase toward the outer regions of the nebulae. On the other hand,
other authors (\cite{balick94}; \cite{gue95}) found element
overabundance high factors (2 to 5) for a few PNe. Particularly,
evidence was found for overabundance of nitrogen from studying the
LISs of some PNe, compared ith the N/H of the main (rims, attached
shells, halos) of the same PNe (\cite{balick94}; \cite{G03}). However,
at least for NGC~7009 (\cite{balick94}; \cite{G03}), the supposed
nitrogen overabundance in LISs was subsequently excluded by a detailed
3D photoionisation modelling of the nebula (\cite{gon06}). Therefore,
nitrogen abundance variations are not reliable, when they are based on
optical data alone, either using long-slit or IFU spectra with their
correspondent ICFs. The main reason for this is that most of the
emission lines from ions of N are not in the optical, but in the UV
(see for example \cite{AB97}; \cite{GV98}; \cite{sta02};
\cite{Henry04}; \cite{gon06}, \cite{gonrev13}, among others). It is
clear, therefore, that the current ionisation-correction methods for
nitrogen are problematic. Moreover, for different reasons, but leading
to a similar situation of low accuracy in the results, the variation
of the sulphur abundances cannot be addressed by studies in the
optical, or optical plus IR either (\cite{Henry04}; \cite{shaw10};
\cite{Henry12}; \cite{gonrev13}, and references therein). This is
because the sulphur anomaly, which refers to the fact that PN sulphur
abundances are systematically lower than those found in most other
interstellar abundance probes for the same O abundance (or
metallicity; \cite{Henry04}). Here again the ICFs seem to be the
problem.  The considerations above indicate that if we aim to address
the possibility of chemical inhomogeneities in PNe, better ICFs are
needed.

The link between different mass-loss episodes (of the PN and previous
phases of the central star) and the resulting chemical inhomogeneities
in the nebula have been investigated in the past, as can be seen in
\cite{balick94}, \cite{hajian97}, \cite{pc98}, \cite{G03},
\cite{goncalves04}, \cite{goncalves09}, and \cite{leal11}, among
others. However, such a link has not been found yet. In the particular
case of PNe with LIS, the fact that these structures are prominent in
the low-ionisation species of elements like N, O, and S, but
particularly, in nitrogen, was tentatively interpreted as indicating
that an overabundance of the latter would account for these
small-scale structures. Interestingly, a number of the LISs studied
(\cite{goncalves01}) show higher velocities than the large-scale
structures they are embedded in, which could indicate a mechanism that
takes enriched material to the outer layers of the nebula. However,
this overabundance has not been found to date. Our results, especially
the N and S abundances, cannot be used to confirm or exclude this
hypothesis because of the problems described previously, although the
lack of strong variations in He and O may indicate that uniform
abundances can be expected throughout mass-loss events that form and
shape PNe.

Concluding, we showed the potential of integral-field observations for
studing extended nebulae compared with the common long-slit
techniques.  On one hand, the agreement found with other literature
studies performed using long-slit spectroscopy only indicates that for
general purpose studies, the commonly employed abundance procedures
are adequate, apart from the ICF problems pointed out.  At the same
time, if detailed spatial resolution is needed, our results highlight
the need for high S/N and high spectral coverage if the
physico-chemical properties are to be computed reliably on a
pixel-to-pixel basis.

\begin{acknowledgements}
  H. Monteiro would like to thank for CNPq grant 573648/2008-5 and
  FAPEMIG grants Grant APQ-02030-10 and CEX-PPM-00235-12 as well as
  INCT-A grant for financial support. DRG acknowledges the partial
  support of FAPERJ (E-26/111.817/2012) and INCT-A as well. MLLF would
  like to thank the Deutscher Akademischer Austausch Dienst
  (DAAD). RLMC acknowledges funding from the Spanish AYA2007-66804 and
  AYA2012-35330 grants. RLMC acknowledges funding from the Spanish
  AYA2012-35330 grant.
\end{acknowledgements}

\end{document}